\documentclass[aps,prb,
twocolumn, 
               a4paper,amsmath,amssymb,superscriptaddress]{revtex4-1}

\usepackage[utf8]{inputenc}
\usepackage[english]{babel}
\usepackage{graphicx}
\usepackage{feynmp}
\usepackage{bm}
\usepackage{color}

\usepackage{color}
\definecolor{Green}{rgb}{0,0.7,0}

\newcommand{\ET}{$\alpha$-(BEDT-TTF)$_2$I$_3$}

\newcommand{\bk}{\bm{k}}
\newcommand{\bkD}{\bm{k}_{\rm D}}
\newcommand{\eD}{\epsilon_{\rm D}}
\newcommand{\ep}{\epsilon }

\newcommand{\g}{\gamma }
\newcommand{\bq}{\bm{q}}

\begin{document}

\title{
Seebeck coefficient of  two-dimensional Dirac electrons 
in an organic conductor under pressure}

\date{Received 29 October 2022; revised 22 April 2023; accepted 25 April 2023; published 9 May 2023}

\author{Yoshikazu \surname{Suzumura}}
 \affiliation{Department of Physics, Nagoya University,
             Nagoya 464-8602, Japan}  
\author{Masao \surname{Ogata}}
 \affiliation{Department of Physics and Trans-scale Quantum Science Institute, University of Tokyo, Bunkyo, Tokyo 113-0033, Japan}
\email{ogata@phys.s.u-tokyo.ac.jp} 

\begin{abstract}
The Seebeck coefficient, which is proportional to a ratio of the thermoelectric conductivity to electrical conductivity has been examined for Dirac electrons in the organic conductor  $\alpha$-(BEDT-TTF)$_2$I$_3$  [BEDT-TTF denotes a molecule given by  bis(ethylenedithio)tetrathiafulvalene] under a uniaxial pressure using  a two-dimensional tight-binding model with both impurity and electron--phonon (e--p) scatterings.  
 We calculate an anomalous temperature ($T$) dependence of the Seebeck coefficient $S_{\nu}$ with $\nu = x$ (perpendicular to the molecular stacking axis) and $y$, which  shows   $S_\nu > 0$  with a maximum at high temperatures and $S_{\nu} < 0$  with a minimum at low temperatures.  
The microscopic mechanism of such a sign change of $S_\nu$ is clarified in terms of the spectral conductivity.
The result is compared with  experiments on $\alpha$-(BEDT-TTF)$_2$I$_3$.
\end{abstract}


\maketitle

\section{Introduction}
 The two-dimensional massless Dirac fermions,\cite{Novoselov2005_Nature438} which show a linear spectrum around the Dirac point have been studied extensively.
 Several properties as a bulk material are explored in an organic conductor~\cite{Kajita_JPSJ2014} given by  \ET ,  where BEDT-TTF denotes bis(ethylenedithio)tetrathiafulvalene.\cite{Mori1984}
 The conductor exhibits a zero-gap state (ZGS)\cite{Katayama2006_JPSJ75} and  the transport property is characterized by the density of states (DOS), which vanishes linearly at the Fermi energy.\cite{Kobayashi2004} 
 The explicit band structure of the Dirac cone is obtained  using  a tight-binding (TB) model, where transfer energies under  pressures are estimated from the extended H\"uckel method.\cite{Kondo2005,Kondo2009} 
The Dirac cone was verified by the first-principles density functional theory (DFT) calculation.\cite{Kino2006} 
Further a two-band model\cite{Kobayashi2007,Goerbig2008} has been proposed  to examine Dirac electrons in an organic conductor. 

Characteristic properties of the Dirac cone appear in the temperature ($T$) dependence of physical quantities. Magnetic susceptibility with a $T$ linear behavior at low temperatures shows a good correspondence between the theory and experiment.\cite{Katayama_EPJ,TakanoJPSJ2010,Hirata2016}
 The chemical potential $\mu$, which also depends on $T$, takes a significant  role in the transport. The reversal of the sign of the Hall coefficient occurs when  $\mu$ becomes equal to the energy of the Dirac point.  
 Such  a sign reversal of the Hall coefficient was proposed theoretically by assuming  the extremely small amount of electron doping.\cite{Kobayashi2008} 
  The sign reversal was also observed  experimentally in the Hall conductivity of \ET.\cite{Tajima_Hall2012} 

Since the conductivity of Dirac electrons is fundamental as the transport, a two-band model with the conduction and valence bands has been studied, where the static conductivity at absolute zero temperature remains finite with a universal value,i.e., independent of the magnitude of impurity scattering owing to a quantum effect.\cite{Ando1998} 
 At absolute zero temperature, the tilting of the  Dirac cone provides the anisotropic conductivity  and  the deviation of the current from the applied electric field.\cite{Suzumura_JPSJ_2014}
At finite temperatures, the conductivity  depends on the magnitude of the impurity scattering, $\Gamma$, which is proportional to the inverse of the life-time by the disorder. 
With increasing temperature, the conductivity increases for $\Gamma \ll T$.\cite{Neto2006} Although  a  monotonic increase in the conductivity is expected for such a model, the measured conductivity (or resistivity) on the above organic conductor shows an almost constant behavior at high temperatures.\cite{Kajita1992,Tajima2000,Tajima2002,Tajima2007,Liu2016}
This is a noticeable transport of the Dirac electron in the presence of the electron-phonon (e-p) interaction, since the  resistivity of the conventional metal at high temperatures increases linearly with respect to $T$  due to the e-p scattering. 
  The   resistivity showing  a nearly constant behavior at high temperatures is explained by  the  acoustic phonon scatterings using  a simple two-band model of the Dirac cone  without tilting.\cite{Suzumura_PRB_2018}
 Although the effect of the e-p scattering at high temperatures is qualitatively understood, the  model should be  improved  to explain  the conductivity  of    the actual organic conductor, where the energy band shows deviation  from the linear spectrum.\cite{Katayama2006_cond}  
   Thus, the TB model with  transfer energies of \ET \; is examined  to show  that the presence of acoustic phonons gives rise to conductivity being nearly constant at high temperatures.~\cite{Suzumura_JPSJ2021}

In addition to the electric conductivity, it is of interest to examine  the thermoelectric (i.e., Seebeck)  effect on the above model, where the $T$ dependence of $\mu$ takes a crucial role. 
The Seebeck coefficient can be obtained microscopically in terms of linear response theory.\cite{Kubo_1957,Luttinger1964}
 However, we have to be careful in treating the heat current, because there are several forms of the heat current depending on the Hamiltonian.\cite{OgataFukuyama}
In the case with impurity potentials and electron-phonon interactions, 
Jonson and Mahan\cite{JonsonMahan} showed that the heat current 
$\bm J_Q$ can be expressed as
\begin{equation}
\bm J_Q = \bm J_Q^{\rm kin} + \bm J_Q^{\rm pot} + \bm J_Q^{\rm e-p(I)}  
+ \bm J_Q^{\rm e-p(II)} + \bm J_Q^{\rm ph},
\label{eq:JM}
\end{equation}
(see, also Ref.\cite{OgataFukuyama}), where $\bm J_Q^{\rm kin}$, $\bm J_Q^{\rm pot}$, and $\bm J_Q^{\rm ph}$ represent the heat current operators originating from the kinetic energy of electrons, from the (impurity) potentials, and from the phonon Hamiltonian, respectively.  
The heat current due to the electron-phonon interaction, $\bm J_Q^{\rm e-p}$, is divided into two contributions. 
If one takes account of only $\bm J_Q^{\rm kin} + \bm J_Q^{\rm pot} + \bm J_Q^{\rm e-p(I)}$ as the heat current operator, one can show that Eqs.~(\ref{eq:L11}) and (\ref{eq:L12}) below (called the Sommerfeld-Bethe relation) hold.\cite{OgataFukuyama}
However, $\bm J_Q^{\rm e-p(II)}$ and $\bm J_Q^{\rm ph}$ do not satisfy the Sommerfeld-Bethe relation and will give additional contributions in the electrothermal conductivity.\cite{JonsonMahan,OgataFukuyama} 
For example, $\bm J_Q^{\rm ph}$ leads to the phonon drag effect.\cite{OgataFukuyama,Matsuura,Matsuura2} 
Jonson and Mahan also discussed that the contribution from $\bm J_Q^{\rm e-p(II)}$ is small in nearly free electron systems. Thus, we do not consider this term and, in the following, we use the Sommerfeld-Bethe relation leaving the phonon-drag problem as a future problem.

So far, there are several theoretical studies 
 on the Seebeck (and Nernst) effect 
in the Dirac electron systems,
\cite{Igor,a014,Kobayashi_2020,a06,a07}
where the Seebeck coefficient exhibits the variety of the sign.
In this paper, we study the Seebeck coefficient for the ZGS 
 of Dirac electrons in the two-dimensional  organic conductor,  \ET.
There have been several experimental  and theoretical studies on this material. 
 As for the experiments, the ZGS has been obtained under both uniaxial pressures  above $P_a = 5$ kbar and hydrostatic pressures above 1.5 GPa.~\cite{Kajita_JPSJ2014}
 Under the uniaxial pressures, the ZGS was found only for $P_a$  corresponding to the pressure along the $a$ direction.~\cite{Tajima2002}
There are several measurements of resistivity suggesting the ZGS under the hydrostatic pressures.~\cite{Kajita1992,Tajima2000,Liu2016}
 Regarding the Seebeck coefficient, the measurement has been performed only for hydrostatic pressures,~\cite{Tajima_Seebeck,Konoike2013} where the sign change of the Seebeck coefficient with decreasing temperature occurs along the $b$ direction.~\cite{Tajima_Seebeck}
However, another experiment\cite{Konoike2013} exhibits the positive  Seebeck coefficient  without the sign change. 
 It could be  ascribed to  the effect of the hole doping, since the latter material is a different sample from the former one. 

As for the theory, there is a work discussing the sign reversal for the Seebeck coefficient of \ET \;   under the hydrostatic pressure.~\cite{Kobayashi_2020}
 However, the sign of the Seebeck coefficient for the $b$ direction 
 obtained in this theory disagrees with that of the experiment.~\cite{Tajima_Seebeck}
 This issue remains as a future problem. In the present paper, 
 we examine the Seebeck coefficient for uniaxial pressures, 
 although the experiment has not yet been performed. 
 We will show the sign change of the Seebeck coefficient in this case.

The present paper is organized as follows. 
 First,  the model and formulation to calculate  the Seebeck coefficient  for \ET \;  with 3/4-filled band are given. 
 Next, after calculating the $T$ dependence of  the chemical potential, we show  the Seebeck coefficient  with the electric conductivity, which  is analyzed in terms of the spectral conductivity.
Finally,  discussions, summary, and  comparison with the experiment  are given.

\section{Formulation}
\begin{figure}
  \centering
\includegraphics[width=6cm]{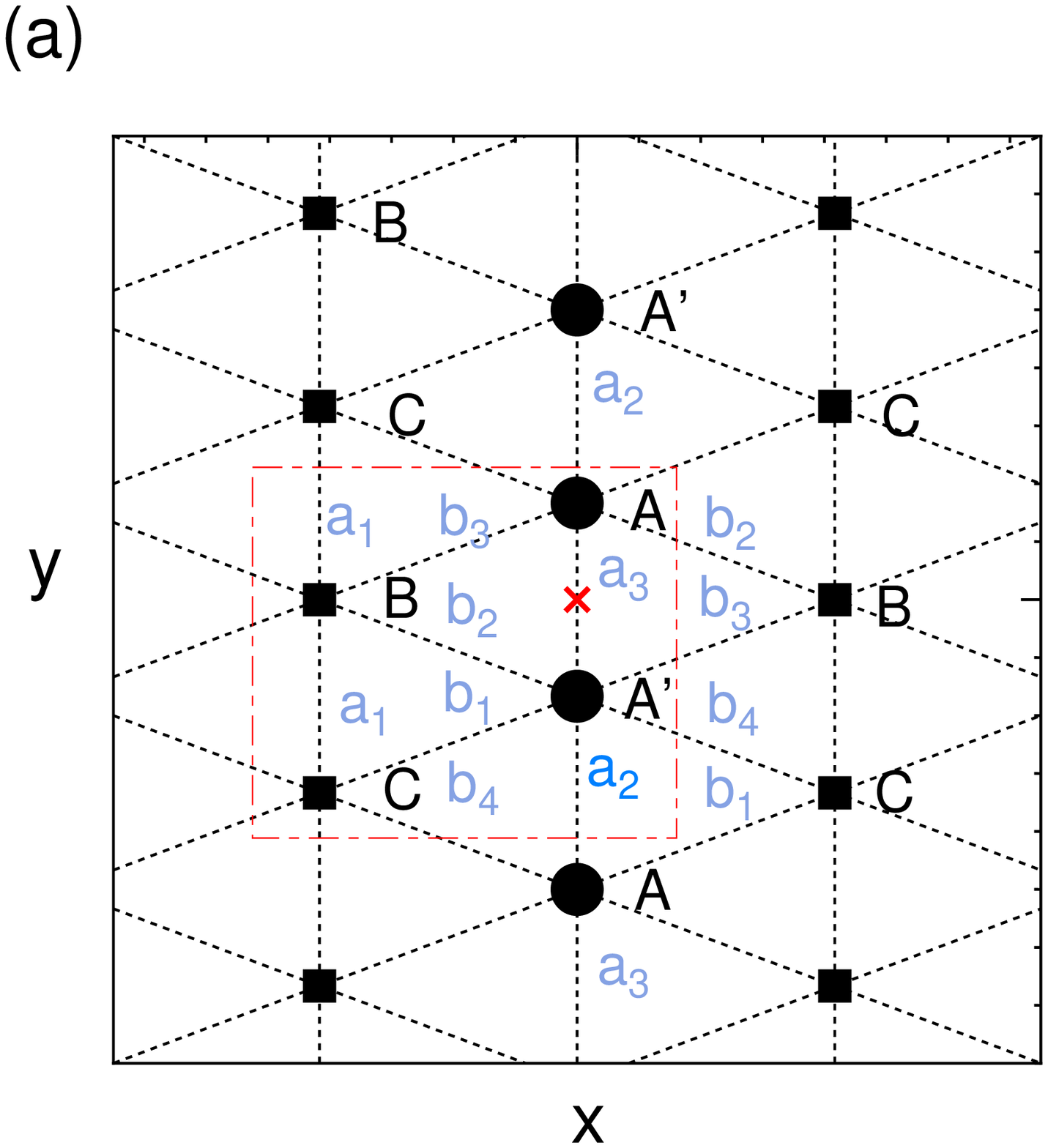}
\includegraphics[width=10cm]{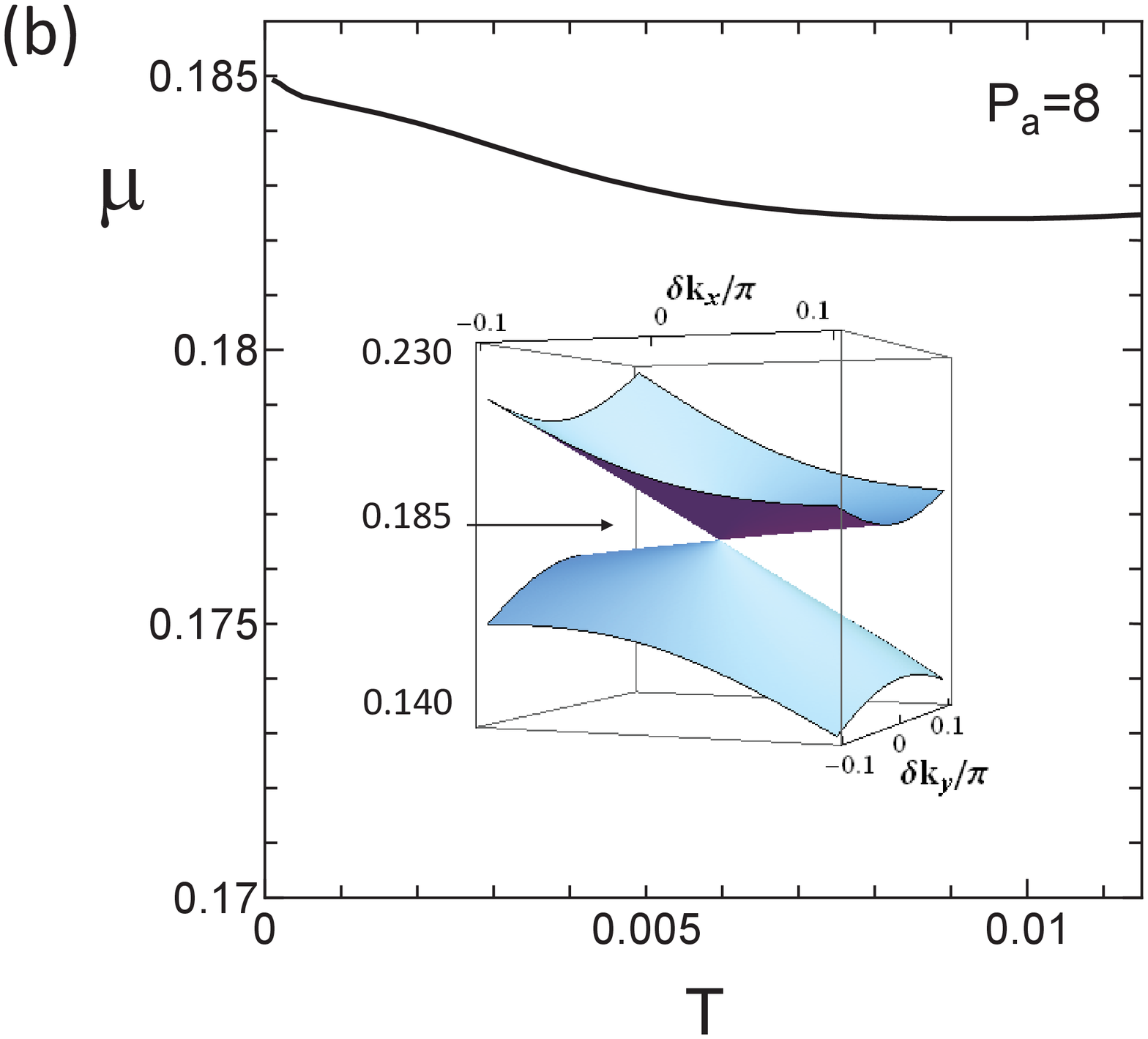}
     \caption{(
(a) Crystal structure, where 
there are four BEDT-TTF molecules 
A, A', B and C in the unit cell indicated by red lines, 
which forms a square lattice.
 Note that $x$ ($y$) corresponds to the $b$ ($a$) direction, which 
 are perpendicular (parallel) to the molecular stacking axis. 
Seven transfer energies 
are shown  by $a_1, \cdots, a_3, b_1 \cdots b_4$ 
 for the nearest neighbor (NN) sites. 
The cross denotes an inversion center between two equivalent molecules 
  A and A'.
(b) Temperature ($T$) dependence of chemical potential ($\mu$) 
 at a uniaxial pressure $P_a$ = 8 kbar.
 The unit is taken as eV. 
The inset denotes a pair of Dirac  cones around a Dirac point, 
 $\bkD = (0.55, 0.25)\pi$, where 
 $(\delta k_x, \delta k_y) = \bk - \bkD$ 
  with  the lattice constant taken as unity.  
 The conduction and valence bands [$E_1(\bk)$ and  $E_2(\bk)$] touch at 
$\bkD$ 
 with a band energy  $\mu_0 = 0.185$ 
corresponding to  the chemical potential at $T$=0.
}
\label{fig1}
\end{figure}

We consider a two-dimensional Dirac electron system per spin,
 which is given by 
\begin{equation}
H = H_0  + H_{\rm p} +  H_{\rm e-p} +H_{\rm imp} \; . 
\label{eq:H}
\end{equation}
 $H_0$ describes a TB model  of 
 the  organic conductor, \ET.
  $H_{\rm p}$ and 
 $H_{\rm e-p}$ describe   an acoustic phonon and 
an  electron-phonon (e--p) interaction, respectively. 
$H_{\rm imp}$ is the impurity potential. 
The unit of the energy is taken as eV.
Figure \ref{fig1}(a) shows  the TB model for $H_0$  
 consisting of four BEDT-TTF molecules in the unit cell. 
 $H_0$ is expressed as 
\begin{eqnarray}
H_0 &=& \sum_{i,j = 1}^N \sum_{\alpha, \beta = 1}^4
 t_{i,j; \alpha,\beta} a^{\dagger}_{i,\alpha} a_{j, \beta} 
                          \nonumber \\
&=& \sum_{\bk}  \sum_{\alpha, \beta = 1}^4
 h_{\alpha \beta}(\bk)  a^{\dagger}_{\alpha}(\bk) a_{\beta}(\bk) 
\; , 
\label{eq:Hij}
\end{eqnarray}
where   $a^{\dagger}_{i, \alpha}$ denotes a creation operator 
 of an electron 
 of molecule $\alpha$ 
 [= A(1), A'(2), B(3), and C(4)]  in the unit cell 
  at  the $i$-th lattice site.  
 $N$ is the total number of square lattice sites and 
 $t_{i,j; \alpha,\beta}$  
 denote the seven kinds of 
 transfer energies $a_1, \cdots, a_3, b_1 \cdots, b_4$
 between  the nearest--neighbor (NN) sites as shown in Fig.~\ref{fig1}(a).
A Fourier transform for the operator $a_{j,\alpha}$ 
 is given by   
 $a_{j,\alpha} = 1/N^{1/2} \sum_{\bk} a_{\alpha}(\bk) \exp[ i \bk \cdot \bm{r}_j]$, 
where $\bk = (k_x,k_y)$ and the lattice constant is taken as unity.
 $H_0$  is diagonalized by  
\begin{eqnarray}
\label{eq:eigen_eq}
\sum_{\beta} h_{\alpha \beta}(\bk) d_{\beta \g(\bk)}
   &=& E_{\g}(\bk) d_{\alpha \g} (\bk)  \; , 
\end{eqnarray}
where $E_1(\bk) > E_2(\bk) > E_3(\bk) > E_4(\bk)$.

The Dirac point ($\bkD$) is calculated  from  
\begin{eqnarray}
\label{eq:ZGS}
E_1(\bkD) = E_2(\bkD)= \eD \; .
\end{eqnarray}
 The ZGS is obtained when 
 $\eD$ becomes equal to  the chemical potential at $T = 0$. 
The chemical potential $\mu$ is determined 
 from the three-quarter-filled condition, which is given by 
\begin{eqnarray}
  \frac{1}{N} \sum_{\bk} \sum_{\gamma}  f(E_{\gamma}(\bk))=   3 \; ,  
  \label{eq:mu}
\end{eqnarray}
where  
 $f(\ep)= 1/(\exp[(\ep -\mu)/k_{\rm B}T]+1)$ with $T$ being temperature 
 and a Boltzmann constant taken as $k_{\rm B }= 1$. 
 Using    the band energy $E_{\gamma}(\bk)$, 
 the  $T$ dependence of $\mu$ is examined 
     in the next paragraph.

On the basis of four molecules in the unit cell of Fig.~\ref{fig1}(a), 
 the matrix element of $h_{\alpha \beta}$ in Eq.~(\ref{eq:Hij})
   is expressed   as
$h_{12}(\bk) = a_3 + a_2 Y $, 
$h_{13}(\bk) = b_3 + b_2 X$,
$h_{14}(\bk) = b_4 Y + b_1 XY$, 
$h_{23}(\bk) = b_2 + b_3 X$, 
$h_{24}(\bk) = b_1 + b_4 X$,
$h_{34}(\bk) = 2 a_1$,  
$h_{11}=h_{22}= h_{33} = h_{44} = 0$ and
$h_{\alpha \beta}(\bk) = h_{\beta \alpha}^*(\bk)$, 
 where   
$X=\exp[i k_x] = \bar{X}^*$, and  $Y= \exp[i k_y] = \bar{Y}^*$. 
Although this model is complicated, when we use these transfer 
energies, we find the zero-gap state (ZGS) composed of Dirac electrons
as shown in the inset of Fig.~1(b), which  consistently explains several 
experimental results. 
 For the uniaxial pressure $P_a$ (kbar), which is applied to 
 the $a$ direction,  transfer energies with NN  sites, 
 $t = a_1, \cdots, b_4$ (eV),   are   estimated
 by the extended H\"uckel method based on  
  the crystal structure analyses  with   the x ray diffraction  measurement. 
 Using  the overlap integrals estimated from  
  the coordinates of the BEDT-TTF molecules,
   transfer energies are obtained   by  
   multiplying  $-10$ eV corresponding to  the atomic energy.
\cite{Mori1984,Kondo2005}
 From   an interpolation method 
 between $P_a$ = 0 and 2 kbar,   
\cite{Kondo2005,Katayama2006_JPSJ75}
 the transfer energies are  given by 
$t(P_a) = t(0) (1+K_t P_a)$, where  $t(0) = a_1(0), \cdots, b_4(0)$ = 
 $ - 0.028$, $ - 0.048$, 0.020, 0.123, 0.140, 0.062, and 0.025,
 and  
 $K_t = $ 0.089, 0.167, -0.025, 0, 0.011, and 0.032, respectively. 
Note that 
 the ZGS is obtained for $P_a > 3$ kbar.

In Fig.~\ref{fig1}(b),
 the chemical potential $\mu$ is shown  as a function of $T$ 
 with a fixed $P_a$ = 8 kbar,  which decreases 
 with increasing for $T (< 0.01)$.
At $T$= 0,  the chemical potential is given by 
 $\mu_0$ = 0.185,  resulting in  the ZGS  as shown 
 on the plane of $\delta \bk = \bk - \bkD$ (the inset), where  
the conduction and valence bands
   touch  at a Dirac point   $\bkD = (0.55, 0.25)\pi$. 
The Dirac cone is  tilted almost along the $k_x$ axis, which gives rise 
 to an anisotropy of the transport property.
With increasing $T$, $\mu$  decreases  
 and takes a slight minimum $\mu$ = 0.1824 at $T \simeq 0.01$. 
 The decrease of $\mu$ suggests that  
 the hole exists in the valence band below the Dirac point.  
The choice of $P_a$ = 8 kbar is large as the extrapolation, but 
  could be used considering the following facts. 
  The Dirac point with increasing $P_a$  is robust due to 
   a small  variation of $\bkD$  compared with the distance from 
   the $\Gamma$ point ($\bk = 0$), where a pair of Dirac points merges at 
    $P _a \simeq 40$  kbar.~\cite{Kajita_JPSJ2014} 
 Furthermore, the ZGS has been observed  up to $P_a$ = 10 kbar
 in the experiment of the resistivity.\cite{Tajima2002}

 In Eq.~(\ref{eq:H}), the second term denotes   the harmonic phonon  given by  
 $H_{\rm p}= \sum_{\bq} \omega_{\bq} b_{\bq}^{\dagger} b_{\bq}$ 
 with $\omega_{\bq} = v_s |\bq|$ and  $\hbar$ =1. 
 The third term is  the e--p interaction 
with a coupling constant $g_{\bq}$, where~\cite{Frohlich}. 
\begin{equation}
 H_{\rm e-p} = \sum_{\bk, \g} \sum_{\bq}
   g_{\bq} c_\g(\bk + \bq)^\dagger c_{\g}(\bk) 
(b_{\bq} + b_{-\bq}^{\dagger})
\; ,
\label{eq:H_e--p}
\end{equation}
 with  $c_{\g}(\bk) = \sum_{\alpha} d_{\alpha \g}a_{\alpha}(\bk)$. 
 The e--p scattering is considered  within  the same band (i.e., intraband) 
 owing  to the energy conservation with $v \gg v_s$, where  $v \simeq 0.05$~\cite{Katayama_EPJ}  denotes the averaged velocity of the Dirac cone. The last 
term of Eq.~(\ref{eq:H}), $H_{\rm imp}$, denotes a normal  impurity scattering.

The spectral conductivity $\sigma_{\nu}(\epsilon,T)$ 
 with $\nu = x$ and  $y$ is calculated as  
\begin{eqnarray}
\sigma_{\nu}(\epsilon,T) &=&  
  \frac{e^2 }{\pi \hbar N} 
  \sum_{\bk} \sum_{\gamma, \gamma'} 
  v^\nu_{\gamma \gamma'}(\bk)^* 
  v^{\nu}_{\gamma' \gamma}(\bk) \nonumber \\
    \nonumber \\
   &  \times &
     \frac{\Gamma_\g}{(\ep - E_{\gamma}(\bk))^2 + \Gamma_\g^2} \times 
 \frac{\Gamma_{\g'}}{(\ep - E_{\gamma'}(\bk))^2 +  \Gamma_{\g'}^2}
  \; ,  \nonumber \\
  \label{eq:spectral}
\\
  v^{\nu}_{\gamma \gamma'}(\bk)& = & \sum_{\alpha \beta}
 d_{\alpha \gamma}(\bk)^* 
   \frac{\partial h_{\alpha \beta}}{\partial k_{\nu}}
 d_{\beta \gamma'}(\bk) \; ,
  \label{eq:v}
\end{eqnarray}
 where  
$h = 2 \pi \hbar$ denotes  Planck's constant.
The spectral conductivity depends on $T$ due to the e-p interaction.
In fact,   $\Gamma_\g$ denotes 
 the damping of the electron of the $\g$ band given by 
\begin{eqnarray}
\Gamma_{\g}  = \Gamma + \Gamma_{\rm ph}^{\g} \; ,
\end{eqnarray}
where the first term comes from the impurity scattering 
and the second term corresponding to  the phonon scattering 
 is given by~\cite{Suzumura_PRB_2018,Suzumura_JPSJ2021,Abrikosov} 
\begin{subequations}
\begin{eqnarray}
  \Gamma_{\rm ph}^\g &=& C_0R \times T|\xi_{\g,\bk}|
  \; ,
 \label{eq:eq16a}
        \\ 
R &=& \frac{\lambda}{ \lambda_0}
 \; ,  
 \label{eq:eq16b} 
\end{eqnarray}
 \end{subequations}
 where $\lambda = |g_{\bq}|^2/\omega_{\bq}$, 
    $\xi_{\g,\bk} = E_{\g}(\bk) - \mu$, 
    $C_0 = 6.25\lambda_0/(2\pi v^2)$   and $\lambda_0/2\pi v = 0.1$.  
 $\lambda_0$ corresponds to  $\lambda$ 
  for an organic conductor\cite{Rice,Gutfreund} and 
  $\lambda$ becomes  independent of $|\bq|$  for small $|\bq|$. 
  $R$ is taken as a parameter. 
   We take  $\Gamma$ = 0.0005 and $R$=0.5  
as in the previous paper,~\cite{Suzumura_PRB_2018}
 where  a choice of $R$=0.5 gives a reasonable suppression of the conductivity at high $T$, and $\Gamma$ = 0.0005 corresponds to a weak impurity scattering due to $\Gamma$ being  much smaller than $T$.

In linear response theory, the electric current density 
$\bm{j} =(j_x,j_y)$ is obtained 
 by  the electric field $\bm{E} = (E_x, E_y)$
 and the temperature gradient $\nabla T$, i.e., the $\nu$ (= $x$ and $y$) 
 component  of the current density, is  expressed as   
\begin{eqnarray}
 j_{\nu} = L_{11}^{\nu} E_\nu  - L_{12}^{\nu} \nabla_\nu T/T
                                                \; , 
   \label{eq:j}
\end{eqnarray}
where $L_{11}^{\nu}$ is the electrical conductivity
 $\sigma_\nu$\cite{Katayama2006_cond} 
 and $L_{12}^{\nu}$ is the thermoelectric conductivity.

From  (\ref{eq:j}), the Seebeck coefficient $S_\nu$ is obtained  by 
\begin{eqnarray}
 S_{\nu}(T) &=& \frac{L_{12}^{\nu}}{T L_{11}^{\nu}} 
                                                \; .
   \label{eq:S}
\end{eqnarray}
As discussed in the introductory part, in terms of Eq.~(\ref{eq:spectral}),
we calculate $L_{11}^{\nu}$ and $L_{12}^{\nu}$ from the 
Sommerfeld-Bethe relation, 
\begin{eqnarray}
L_{11}^{\nu} &=&  \sigma_{\nu}(T) = 
              \int_{- \infty}^{\infty} d \ep  
            \left( - \frac{\partial f(\ep) }{\partial \ep} \right)
         \times  \sigma_{\nu}(\epsilon,T)  \; ,
                                     \nonumber \\
   \label{eq:L11}
                                                \\
 L_{12}^{\nu} &=&  \frac{-1}{e}
              \int_{- \infty}^{\infty} d \ep  
  \left( - \frac{\partial f(\ep) }{\partial \ep} \right)
         \times (\ep - \mu) \sigma_{\nu}(\ep,T)
       \; , \nonumber \\
   \label{eq:L12}
\end{eqnarray}
 where  $e (>0)$ denotes  the electric charge.
Noting that $- \partial f(\ep) /\partial \ep$ is the even function 
of $\ep -\mu$, and $\sigma_\nu(\ep,T)$ in 
Eq.~(\ref{eq:L11}) can be expanded as 
\begin{eqnarray}
  \sigma_\nu(\ep,T) &=&  \sigma_\nu(\mu,T)
       + \sigma_{\nu}^{'}(\mu,T)(\ep - \mu)  
                    \nonumber    \\
    && + \frac{1}{2} \sigma_{\nu}^{''}(\mu, T)  (\ep - \mu)^2    + \cdots  ,
      \label{eq:L11_T}       
\end{eqnarray}
Eq.~(\ref{eq:L12}) is calculated as
\begin{eqnarray}
  e L_{12}^{\nu}(T) &=&  
       - \frac{\pi^2}{3} \sigma_{\nu}^{'}(\mu,T)T^2
       - \frac{7\pi^4}{90} \sigma_{\nu}^{'''}(\mu,T)T^4 
                  \nonumber         \\
        &&    + \cdots ,
      \label{eq:L12_T}         
\end{eqnarray}
at low temperatures.It  is shown later that   
 the sign change of $S_\nu(T)$ with decreasing $T$ 
 comes from that of the first term  of Eq.~(\ref{eq:L12_T}). 
 
\begin{figure}
  \centering
\includegraphics[width=8cm]{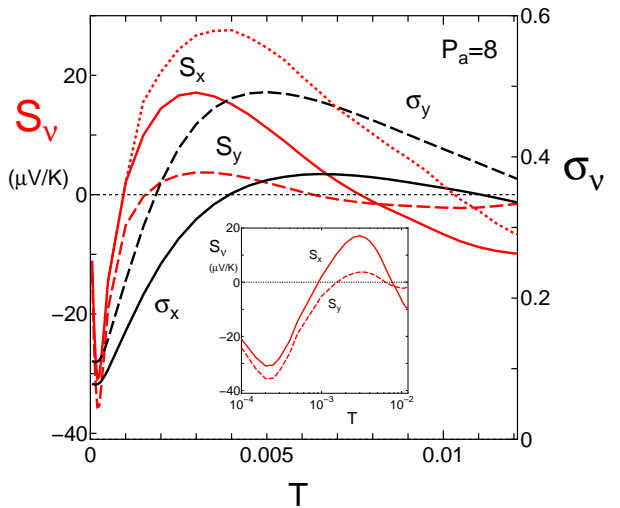}
     \caption{
$T$ dependence of the Seebeck coefficients $S_{x}$ (red solid line) 
 and $S_{y}$ (red dashed line) for $P_a$ = 8 kbar, 
 which are compared with  
  electric conductivities 
 $\sigma_x$ (solid line) and  $\sigma_y$ (dashed line). 
 The e--p coupling  is taken as  $R$=0.5. 
The dotted line shows $S_{x}$ for $R$=0.  
The inset denotes the magnified $S_x$ and $S_y$, 
 which 
 suggest $S_\nu \rightarrow 0$ at $T \rightarrow 0$. 
  }
\label{fig2}
\end{figure}

\begin{figure}
  \centering
\includegraphics[width=8cm]{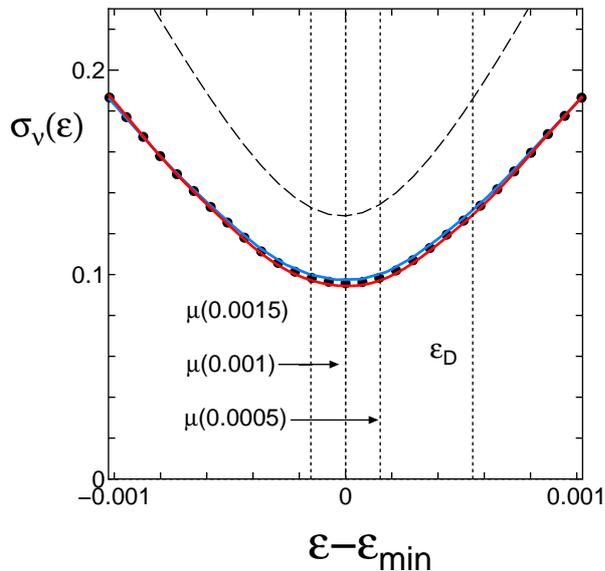}
     \caption{
 Spectral conductivity $\sigma_{\nu}(\epsilon,T)$   for  $\nu= x$  
 as a function of  $\ep - \ep_{\rm min}$, which  are obtained for 
  $T$ = 0.0005 (red line),  0.001 (dots), and 0.0015 (blue line). 
  $\ep_{\rm min} \simeq \mu (0.001)$.
The dashed line shows $\sigma_{y}$  for  $T$ = 0.001.
The vertical lines denote 
   locations of the chemical potential $\mu (T)$   for  
   $ \ep_{\rm D} = \mu (0) \simeq 0.1850$ , 
  $\mu (0.0005) \simeq $  0.1846, 
  $\mu (0.001) \simeq $ 0.1845, and 
  $\mu (0.0015) \simeq $ 0.1843 [(Fig.~\ref{fig1}(b)]. 
The case of  $\mu (T) < \ep_{\rm min} $  gives  $S_{\nu} > 0$, 
 while the case of  $\mu (T)> \ep_{\rm min} $  suggests  $S_{\nu} < 0$.
 }
\label{fig3}
\end{figure}

\section{Seebeck coefficient}
Now we study  $S_{\nu}(T)$  using  parameters of \ET. 
The  Seebeck coefficient  of \ET \; under uniaxial pressures    
provides the following  $T$ dependence.
In the present  paper, we take $P_a$ = 8, which shows a sign change, where  
  $S_x(T) > 0  ( < 0)$ at high temperatures  (at low temperatures). 
Figure \ref{fig2} shows 
 the $T$ dependence of the Seebeck coefficient, $S_{\nu}$ 
 and the electrical conductivity $\sigma_\nu$, 
  where  $\sigma_y > \sigma_x$    for any $T$  
 and  $S_x > S_y$ for $T < 0.008$. 
Note that both $S_x$ and $S_y$ exhibit the change of the  sign 
  at low temperatures. It is found that  $S_{x}=0$  at $T \simeq 0.0009$  
and that $S_{x}$ takes a maximum $\simeq 17 \mu V/K$ at $T \simeq $  0.003. 
At low temperatures given by  $S_x < 0$, 
  $S_x$  takes a minimum. 
Similar behavior is also obtained  for $S_y$, where 
 $S_{y} = 0$  at  $ T \simeq 0.0015$ and 
 the temperatures corresponding to the maximum and minimum 
 are almost the same as those of $S_x$. 
The relation   $\sigma_y > \sigma_x$, which  comes from 
  the tilted Dirac cone [Fig.~\ref{fig1}(b)],~\cite{Suzumura_JPSJ_2014}
 results in $S_x > S_y >0$. 
The inset denotes magnified $S_{\nu}$ at low temperatures. 
A  minimum exists at $T \simeq 0.0002$ 
 and the extrapolation to lower temperatures
 suggests $S_{\nu} \propto - T$ since   
 $S_\nu \sim  - T \sigma_{\nu}^{'}(\mu,T)/\sigma_{\nu}(\mu,T)$ 
 [see  Eqs.~(\ref{eq:L11_T})  and (\ref{eq:L12_T})].
 The interband effect ($\gamma \not= \gamma'$ ) becomes 
 small  at low temperatures and 
 the increase  of $\Gamma$ gives a slight reduction of $S_\nu$.  
 The decrease of the uniaxial pressure $P$ reduces the temperature region 
 for $S_x > 0$. 
 Note that there is enough range of  $P_a$ for 
  a sign change of $S_x$, which is in general sensitive to parameters. 
In fact, $S_x$ for $P_a$ = 6 kbar (not shown here) also shows 
 the sign change at  $T \simeq 0.0005$.

 In order to comprehend the existence of $S_x(T) = 0$, 
 we  examine   the spectral conductivity.
 In Fig.~\ref{fig3}, spectral conductivity $\sigma_{\nu}(\ep,T)$ 
 is shown as a function of $\ep - \ep_{\rm min}$, 
 where $\sigma_{\nu}(\ep,T)$ takes a minimum  
  at $\ep_{\rm min}= 0.18447$. 
 The minimum is close but lower than that of 
  the Dirac point   $\ep_{\rm D}$   ($\ep_{\rm D} - \ep_{\rm min} \simeq 0.0005$). 
 $S_x(T) = 0$ occurs  
when $\mu(T)\simeq \ep_{\rm min}$ at some temperature.
 A similar minimum is obtained for $\sigma_{y}(T)(\ep,0.001)$ (dashed line),
  which is larger than $\sigma_{x}(\ep,0.001)$.  $\sigma_{x}(\ep,T)$ is shown 
 for the fixed $T$ = 0.0005, 0.001 and 0.0015,  where the width depends 
   on $T$ due to $\Gamma_{\rm ph}^\gamma$ (Eq.~{\ref{eq:eq16a}). 
The vertical lines denote the corresponding   $\mu(T)$, 
 where $\mu(0.001)= \ep_{\rm min}$ and  $\ep_{\rm D} = \mu(0)$.
Since  $\mu(0.0015) < \mu(0.001) < \mu(0.0005)$,
 $\sigma_x{'}(T) > 0$ for $T < 0.001$ and 
  $\sigma_x{'}(T) < 0$ for $T > 0.001$.
From Eq.~(\ref{eq:L12_T}), it turns out that 
$S_{\nu} > 0$ is obtained for   $\mu (T) < \ep_{\rm min} $ 
 and  $S_{\nu} < 0$ is obtained for $T \simeq 0.00095 < 0.001$, i.e., 
   for  $\mu (T)$ being slightly lower   than  $\ep_{\rm min} $ 
  due to the  second term  of   Eq.~(\ref{eq:L12_T}).
Thus, with decreasing $T$,  $S_{\nu}(T)$ changes the sign from a positive 
 to a negative  one   at $\mu \simeq  \ep_{\rm min}$ 
   corresponding to $\sigma_\nu^{'}(\mu,T)$ = 0. 
 Note that the sign change of $S_\nu$ 
 in Fig.~\ref{fig2} is obtained  in the case of $\mu < \eD$.
This fact is different from that of the Hall coefficient,~\cite{Kobayashi2008}
 where the sign change occurs at $\mu =  \eD$.  

Here we note  the minimum and maximum of $S_{\nu}$ in Fig.~\ref{fig2}. 
Such a behavior is also obtained only for the impurity 
 scattering, i.e., without the e--p  coupling ($R$=0). 
 Compared with  the dotted line in Fig.~\ref{fig2}, 
  $S_x$ at high temperature is reduced by the e-p coupling, 
 while  $S_x$ at low temperatures ($T <0.001$) remains the same.
We also examined  $S_{\nu}$ at lower pressures.
 For $P_a$=6, it is found  that $S_x$ decreases and $S_y$ increases, while  
 the maximum and minimum still  exist.
The  spectral conductivity $\sigma_{\nu}(\ep)$ shows  
 the existence of  the minimum and 
 the  $T$ dependence of  the chemical potential similar to  Fig.~\ref{fig3}. 
 
\section{Discussions and Summary}
Here, we  discuss the relevance of our result to experiments.
The temperatures of the sign change and the maximum 
of $S_x (> 0)$  in Fig.~\ref{fig2} are  similar to those obtained in the 
experiment under the hydrostatic pressures.\cite{Tajima_Seebeck} 
Although this is suggestive, note that the experiment is carried out 
in the hydrostatic pressure, 
 while our calculation is for the uniaxial pressure. 
As another aspect, a minimum of $S_x$  at low temperature, suggesting 
 $S_x \rightarrow 0$ at  $T = 0$, 
 is an interesting  piece of information from our calculation, 
 that should be examined experimentally  by decreasing the temperature.  

Finally, let us comment on the Seebeck coefficient 
in the case of the hydrostatic pressure. 
 The previous theory\cite{Kobayashi_2020}
 studied the effect of short-range repulsive interactions on 
 the TB model  with  the transfer energies obtained 
      from the first-principles calculation,~\cite{Kino2006}
  and showed that the decrease of $T$ leads to 
 the sign change from $S_y > 0$ into $S_y < 0$   at $T \simeq 0.0002$.
 Noting that the Seebeck coefficients  are  in general sensitive to parameters 
  such as  transfer energies and site potentials, 
   we examined $S_x$ and $S_y$ for   the following  two cases. 
 One is a model used in the previous 
calculation~\cite{Katayama_EPJ,Suzumura_JPSJ2021}
(but slightly different from that used in Ref.\cite{Kobayashi_2020}), 
 in which the transfer energies  obtained 
  from the first-principles calculation~\cite{Kino2006}
    are fixed at a low temperature, 
 and  the site potentials obtained from  the mean field of the interaction  
  are taken as those at  $T$ = 0.  In this case, 
 we obtained $S_y > 0$ at high temperatures followed by the sign change 
 at low temperatures, while   $S_x$ is negative at any temperature. 
The other is a model in which the transfer energies are 
  obtained by crystal structure analyses  at $P$ = 1.76 GPa.~\cite{Kondo2009} 
 Using a choice of site potentials that gives a ZGS,~\cite{Kondo2009} 
   we obtained  that  $S_y >0$ at high temperatures with the sign change 
 at a temperature being slightly higher than that in the former model, 
      while   $S_x$ is negative  at  any temperature.
 Thus, we found $S_x < 0$ as a common feature of the above two models, 
  which  is inconsistent with the  experiment.~\cite{Tajima_Seebeck} 
 It remains a future problem 
 to obtain a reliable TB model exhibiting  the sign change of $S_x$ 
   for hydrostatic pressures. 

In summary, 
 for the $T$ dependence of the Seebeck coefficient of \ET, $S_\nu (T)$ 
 under uniaxial pressures was calculated 
 although there is no experiment at present.  
 We obtained the sign change for both $S_x$ and $S_y$ 
 and clarified the microscopic mechanism in terms of the spectral conductivity
 $\sigma_\nu(\mu,T)$. 
 The  correspondence of the present theory to the experiment  awaits 
   the future  measurement of the Seebeck coefficient under uniaxial pressures.

\acknowledgements
We thank N. Tajima for useful discussions and for providing us with data of the Seebeck coefficient 
of $\alpha$-(BEDT-TTF)$_2$I$_3$ at 1.9 GPa. 
This work is supported by JST-Mirai Program Grant (Grant No. JPMJMI19A1).


\end{document}